\documentclass{llncs}
\usepackage[utf8]{inputenc}
\usepackage{booktabs}
\usepackage{multirow}
\usepackage{siunitx}
\usepackage{caption}
\usepackage{natbib}
\usepackage{graphicx}
\usepackage{pgfplots}
\usepackage{pgfplotstable}
\pgfplotsset{compat=1.7}
\usepackage{tikz}
\usetikzlibrary{positioning}
\usepackage{authblk}

\title{Neural document expansion for ad-hoc information retrieval}
\author{Cheng Tang \footnote{tcheng@amazon.com}, 
Andrew Arnold \footnote{anarnld@amazon.com}}
\institute{Amazon}
\begin{document}

\maketitle
\begin{abstract}
    Recently, \citet{nogueira2019DE} proposed a new approach to document expansion based on a neural Seq2Seq model, showing significant improvement on short text retrieval task. However, this approach needs a large amount of in-domain training data.
    In this paper, we show that this neural document expansion approach can be effectively adapted to standard IR tasks, where labels are scarce and many long documents are present.
\end{abstract}

\section{Introduction}
In traditional ad-hoc retrieval, queries and documents are represented by variants of bag-of-words representations. This leads to the so called vocabulary mismatch problem: when a query contains words that do not exactly match words in a relevant document, the search engine may fail to retrieve this document.
Query expansion and document expansion, the methods of adding additional terms to the original query or document, are two popular solution to alleviate the vocabulary mismatch problem. 

Document expansion has been shown to be particularly effective for short text retrieval and language-model based retrieval \citep{DE_for_LM, efron12}.
Most of the existing works in document expansion are unsupervised:
using information from the corpus to augment document representation, e.g., retrieval based \citep{efron12} and clustering based \citep{croft04, DE_for_LM}, or using external information to augment document representation \citep{DE_with_wordnet, DE_with_external_collection}.

Recently, \citet{nogueira2019DE} proposed a new approach to document expansion, which is based on a popular generative sequence-to-sequence model (Seq2Seq) in NLP, transformers \citep{wolf2020huggingfaces}. It leverages supervision to train the model to predict expansion terms conditional on each document. The paper has shown significant improvement on passage (short document) datasets, when trained in-domain.
In this paper, we follow this line of supervised neural document expansion approach and explore its performance on standard IR benchmarking dataset. Our main contributions are: 1. Adapting the method to unlabeled datasets by exploring transfer learning and weak-supervision approaches.
2. Adapting the method to traditional IR datasets, where a large number of long documents are present.

\begin{figure}[htb]
\begin{minipage}{0.5\textwidth}
\begin{tikzpicture}
  \node (img)  {\includegraphics[scale=0.38]{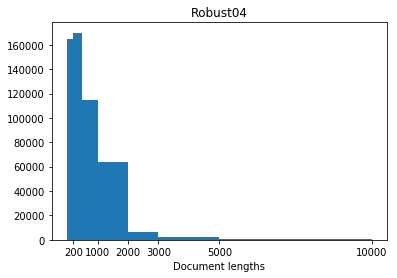}
  };
  \node[left=of img, node distance=0cm, rotate=90, anchor=center,yshift=-0.7cm,font=\color{black}] {document counts};
 \end{tikzpicture}
\end{minipage}%
\begin{minipage}{0.5\textwidth}
\begin{tikzpicture}
  \node (img)  {\includegraphics[scale=0.38]{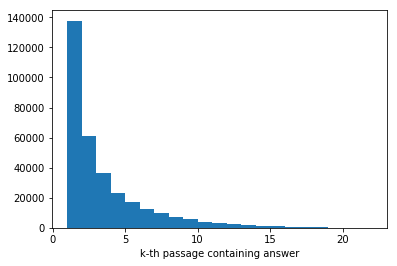}};
\end{tikzpicture}
\end{minipage}%
\caption{
Left: document lengths distribution in Robust04; 
Right: distribution of relevant passage (answer-containing passage) indices in MSMarco-documents
}
\label{fig:histograms}
\end{figure}

\section{Document retrieval with Seq2Seq models}
\label{sec:DE_model}
In this section, we discuss our approach to document expansion.

\paragraph{\textbf{Train a passage level document expansion model}}
We follow the approach of \citet{nogueira2019DE}: we use data of format $(p, q)$, where $q$ is a question and $p$ is a passage \textit{relevant} to it. A Seq2Seq model is trained so that conditional on input $p$, it learns to generate the ground-truth question $q$. The dataset we found most effective for training the document expansion model is the MSMarco passage-ranking dataset \citep{bajaj2018ms}.

\paragraph{\textbf{Domain adaptation}} 
Since most IR datasets are short of annotated queries, it is important to explore ways to adapt the document expansion model to out-of-domain unlabeled datasets. We explored two simple approaches: zero-shot transfer and weak supervision (retrieval-based pseudo annotation).
In zero-shot transfer, we directly apply a model trained from one domain to another; in retrieval-based pseudo annotation, we issue out-of-domain queries to a target corpus and treat the top-$k$ returned documents as ``relevant''.

\paragraph{\textbf{Adaptation to long documents}}
Due to the memory overhead of transformer-based models, the maximal input length of this family of models are constrained to be short.
We trained our expansion models on passage-level datasets. 
However, during inference time, a standard IR dataset is usually a mixture of short and long documents, and expanding on an arbitrarily capped portion of a document may yield sub-optimal performance. Figure~\ref{fig:histograms}-Left shows the document lengths distribution of Robust04 dataset. We see that most of the documents are around 1000-tokens long, which has exceeded the typical input length transformer based models can take \citep{long_text_retrieval}.
To deal with this problem, we explored three strategies (see experiment results at Table \ref{tab:robust04}: ``Long document generation methods'' section):
\begin{enumerate}
    \item \textbf{concatenation of paragraph generation (CONCAT in Table \ref{tab:robust04})}: Split each long document into overlapping passages, respecting natural sentence boundaries (with sliding window size to be around 60 tokens, since this is roughly the median size of passages in MSMarco dataset, which we used for training) and run expansion inference on each passage. All generated expansions are directly appended to the original documents.
    \item \textbf{first k sentences (FIRST-K in Table \ref{tab:robust04})}: Run expansion model on the first $k$ whole sentences of each document. This strategy is based on our analysis of MSMarco datasets. From MSMarco-QA dataset, we can obtain $(q, p, a)$ triplets so that we know passage $p$ contains an answer for question $q$. We treat this as a signal of question-passage relevance. Then we can trace these passages back to the document they belong to using the MSMarco-documents dataset. 
    Comparing ``Capped passage generation methods'' to ``Long document generation methods'' in Table \ref{tab:robust04}, we see that the ``Long document generation methods'', and in particular, the CONCAT and FIRST-K generation methods are more effective.
    \item \textbf{passage-importance indicator (PI in Table \ref{tab:robust04})}: Before seeing any questions, we predict which passage will be queried using a unsupervised predictor \citep{doc_homogeneity_measure}, and generate expansion exclusively based on this passage. In our experiments, this method do not yield good performance, likely due to the low quality of the passage-importance predictor. 
\end{enumerate}


\section{Experiments}
To train the document expansion models, we selected datasets that are typically used for passage-ranking and question-answering tasks, i.e., MSMarco passage-ranking \citep{bajaj2018ms}, Natural Questions \citep{natural_questions}, and SQuAD \citep{squad_v2}. In addition, we also used a TREC collection short-document dataset, microblogs \citep{Lin2014}.
To evaluate the performance of document expansion method on information retrieval, we additionally add the standard IR benchmarking dataset Robust04 \citep{robust_04}.
Our baseline retrieval methods, BM25 and Query Likelihood (QL), use implementation from the open-source Anserini toolkit \citep{anserini}; for Seq2Seq models, we experimented with transformers (OpenNMT implementation \citep{klein-etal-2017-opennmt}) and pre-trained BART (Huggingface Pytorch implementation \citep{wolf2020huggingfaces}). 

\subsection{Retrieval performance on passage datasets}
From Table \ref{tab:in_domain}, we see that retrieval-based weak supervision, while showing good performance as training signal for neural retrieval models \citep{NRM_weak_supervision}, does not yield good document expansion models.
Instead, the BART-based zero-shot transfer model has competitive performance to in-domain trained-from-scratch models. Once decided on the zero-shot transfer learning approach, we tried several fine-tuning strategies with the BART model (Table \ref{tab:train_strategy}), drawing inspiration from \citep{Yilmaz2019}. We found that fine-tuning pre-trained BART with MSMarco-passages dataset and with a mixed MSMarco-passages and microblogs dataset produces the best document expansion models.
Although less effective, our experiment suggests that other passage-level datasets such as Natural Questions and SQuAD, can also be used as sources to train document expansion models.

\begin{center}
  \begin{tabular}{l|SSSSSS}
    \toprule
    \multirow{4}{*}{DE models} &
      \multicolumn{2}{c}{MSMarco-passages} &
      \multicolumn{2}{c}{Natural Questions} &
      \multicolumn{2}{c}{SQuAD-v2} \\
      & {Trec-DL} &  {Trec-DL} \\
      & {(DEV)} & {(DEV)} \\
      & {R@100} & {MAP} & {R@100} & {MAP} & {R@100} & {MAP} \\
    
    \midrule
    {\small BM25} 
    & {0.4531} & {0.3773} & {0.7619} & {0.3616} & {0.9737} & {0.7997} \\
     & {(0.7619)} & {(\textbf{0.3616})} & {} & {} & {} & {} \\
    \midrule
    {\small \textbf{In-domain trained}}
    & {0.4545} & {0.3872} & {\textbf{0.8671}} & {0.4450} & {0.9754} & {0.7915}
    \\
    {\small \textbf{transformer}}
    & {(0.7127)} & {(0.2203)} & {} & {} & {} & {} \\
    
    \midrule 
    {\small \textbf{Weakly supervised}} \\
    {\small \textbf{transformer}} 
    & {NA} & {NA} & {0.7649} & {0.3608} & {0.9717} & {0.7913} \\
    
    \midrule
    {\small \textbf{Zero-shot transfer}} \\
    {\small (transformer)} 
    & {NA} & {NA} & {0.7773} & {0.3879} & {0.9764} & {0.8056} \\
    
    {\small (fine-tuning} 
    & {\textbf{0.5297}} & {\textbf{0.465}} & {0.8302} & {\textbf{0.4501}} & {\textbf{0.9787}} & {\textbf{0.8121}} \\
    {\small BART)}
    & {(\textbf{0.7949})} & {(0.2674)} & {} & {} & {} & {} \\
    \bottomrule
  \end{tabular}
  \captionof{table}{In-domain trained and weakly-supervised document expansion model; for MSMarco-passages, we have two test sets: DEV and Trec-DL \citep{trec_dl_2019}}
  \label{tab:in_domain}
\end{center}

\begin{center}
  \begin{tabular}{l|SSSS|SS}
    \toprule
    \multirow{2}{*}{DE models} &
      \multicolumn{2}{c}{Natural Questions} &
      \multicolumn{2}{c|}{SQuAD-v2} &
      \multicolumn{2}{c}{Robust04} \\
      & {R@100} & {MAP} & {R@100} & {MAP} & {R@100} & {MAP} \\
      \midrule
    {\small BM25} 
     & {0.7619} & {0.3616} & {0.9737} & {0.7997} & {0.4152} & {0.2553} \\
    {\small MSMarco-passages}
     & {\textbf{0.8302}} & {\textbf{0.4501}} & {\textbf{0.9787}} & {0.8121} & {\textbf{0.4229}} & {0.2620} \\
    {\small MSMarco-passage} \\
    {$\rightarrow$ microblogs}
     & {0.7931} & {0.4} & {0.9757} & {0.7962} & {0.4206} & {0.2533} \\
    {\small MSMarco-passages} \\
    {$+$ microblogs} 
     & {0.8239} & {0.4437} & {\textbf{0.9787}} & {\textbf{0.8133}} & {0.4212} & {\textbf{0.2630}} \\
     \cmidrule{1-5}
     {\small Natural Questions} 
      & {\textbf{0.9031}} & {\textbf{0.5271}} & {0.9782} & {0.8099}
     & {0.4190} & {0.2626} \\
     {\small SQuAD-v2}
      & {0.8173} & {0.4228} & {\textbf{0.9798}} & {\textbf{0.8156}}
     & {0.4218} & {0.2616} \\
    \bottomrule
  \end{tabular}
\captionof{table}{Zero-shot transfer of passage-level DE model}
\label{tab:train_strategy}
\end{center}





\subsection{Retrieval performance on Robust04}
To test the performance of passage-level trained document expansion model on standard IR corpus, we fixed the passage-level model to be ``MSMaroc-passage + microblog'' trained. Then we explored three expansion generation methods as described in Section~\ref{sec:DE_model}. The result of applying passage-level trained document expansion model to Robust04 can be found in Table \ref{tab:robust04}. 
\paragraph{\textbf{Traditional IR baselines}}
In addition to tf-idf based BM25, we applied document expansion to two popular language-model based retrieval methods: query-likelihood with Dirichlet Smooth (QLD) and query-likelihood with Jelinek-Mercer smoothing (QLJM). 
We found that document expansion has good performance with QLD. We speculate that this is because the way our current document expansion model works is similar to the Dirichlet smoothing model, which models the process of adding unseen words to the original documents by drawing from a background distribution based on the entire corpus. Here, the document expansion model in addition samples words from the query distribution (conditional on each document).
Since document expansion does not significantly improve QLJM, we did not include it in the rest of our experiments.
\paragraph{\textbf{Capped passage generation vs Long document generation}}
Comparing ``Capped passage generation methods'' to ``Long document generation methods'', we see that for both BM25 and QLD, the two long document expansion methods, CONCAT and FIRST-K, have better performance.
The fact the CONCAT has the best performance suggests that even if most queries are targeting the head portion of a document (Figure \ref{fig:histograms}-Right), using information from the entire document may still be beneficial to expansion generation.
\paragraph{\textbf{Performance with pseudo-relevance feedback and BERT re-ranker}}
Since document expansion is one of several techniques that can improve the document retrieval performance, we also want to understand how it works when combined with other techniques. In our experiments, we first explored combining the best performing document expansion model with RM3 \citep{rm3}, a popular query expansion method based on pseudo-relevance feedback. While RM3 alone significantly improves the baseline retrieval performance, DE** \footnote{DE** indicates concatenation of paragraph generation with MSMarco+microblog trained passage model.} can still add an additional boost on top. 
We want to point out that comparing to RM3, which requires two rounds of retrieval at query time, document expansion model is applied offline and does not add additional computational overhead at run time. 
BM25 with document and query expansion makes a first-stage ranker in the ranking pipeline. In our last experiment, we test its end-to-end performance when combined with a second-stage neural ranker, BERT \citep{nogueira2019passage}. To evaluate the end-to-end result, we used metrics $R@k$ and $P@k$ for small $k$, mimicking what the ranking system presents to a user (the top-$k$ ranked documents). 
Our experiment results indicate that our document expansion models are complementary to query expansion as a first-stage ranker and can improve the end-to-end ranking performance when combined with a second-stage ranker.

\begin{center}
  \begin{tabular}{l|SSSS}
    \toprule
    \multirow{3}{*}{Experiment category} &
    {Methods} &
      \multicolumn{2}{c}{Robust04} \\
      & {} & {R@100} & {MAP}  \\
      \midrule
     {Traditional}
     &
     {\small BM25} 
     & {0.4152} &{0.2553} \\
     {IR baselines} &
     {\small QLD}
     & {0.4157} &{0.2502} \\
     &
     {\small QLJM}
     & {0.3995} &{0.2287} \\
     \midrule
     {Capped passage}
     &
     {\small BM25+passage-DE*} 
     & {0.4212} &  {\textbf{0.2630}} \\
     {generation methods} & 
     {\small QLD+passage-DE*} 
     & {\textbf{0.4270}} &  {0.2620} \\
     &
     {\small QLJM+passage-DE*} 
     & {0.4058} &  {0.2350} \\
     
     
    \midrule
    {Long document}
    &
    {\small BM25+DE* (CONCAT)} 
    & {\textbf{0.4283}} & {\textbf{0.2631}} \\
    {generation methods}
     
    &
    {\small BM25+DE* (FIRST-K)} 
    & {0.4226} & {0.2625} \\
    &
    {\small BM25+DE* (PI)} 
    & {0.4212} & {0.2588} \\
     \cmidrule{2-4}
    & 
    {\small QLD+DE* (CONCAT)} 
    & {\textbf{0.4290}} & {0.2615} \\
    &
    {\small QLD+DE* (FIRST-K)} 
    & {0.4272} & {\textbf{0.2625}} \\
    &
    {\small QLD+DE* (PI)} 
    & {0.4259} & {0.2577} \\
    
    \midrule
   {DE+pseudo-relevance feedback}
   &
   {\small BM25+RM3}
   & {0.4517} & {0.2941} \\
   &
   {\small BM25+RM3+DE**}
    & {\textbf{0.4641}} & {\textbf{0.3035}} \\
   %
   \midrule
   \multirow{3}{*}{DE + BERT reranker}\\ &
   \multicolumn{3}{c}{End-to-end metrics} \\
      & {R@10} & {P@10}  & {P@5}\\
   \cmidrule{1-4}
   {\small BM25 + BERT}
    & {0.2771} & {0.3731} & {0.4137}\\
   {\small BM25 + RM3 + BERT}
    & {0.2608} & {0.3803} & {0.4048} \\
   {\small BM25 + DE** + BERT} 
    & {\textbf{0.2824}} & { 0.3767} & {0.4161} \\
   {\small BM25 + RM3 + DE** + BERT}
    & {0.2726} & {\textbf{0.3944}} & {\textbf{0.4241}} \\
    \bottomrule
  \end{tabular}
\captionof{table}{Robust04 experiments (DE*: MSMarco-passage+microblog trained passage model; DE**: concatenation of paragraph generation, CONCAT, with DE*)}
\label{tab:robust04}
\end{center}

\section{Conclusion}
We showed that a document expansion model trained on passage-level datasets of (question, relevant passage) pairs can be directly applied to out-of-domain IR datasets to improve retrieval performance. 
We explored simple approaches to adapt the model to standard IR dataset (Robust04) where a large number of long documents are present, and we showed that adapting the passage-level trained model to long documents further improves retrieval performance.
However, our current simple adaptations to long documents do not significantly improve the model performance (see Table \ref{tab:robust04}, Long document generation methods). We cannot conclude whether this is due to the nature of the relevant passage distribution over long documents (i.e., they tend to be the first few passages of any document, according to Figure \ref{fig:histograms}-Right). Hence, it may be worth exploring model architectures that allow longer input sequences, for example, by switching to sparse attention layers \citep{beltagy2020longformer}.

\bibliography{mybibs}

\begin{thebibliography}{22}
\providecommand{\natexlab}[1]{#1}
\providecommand{\url}[1]{\texttt{#1}}
\expandafter\ifx\csname urlstyle\endcsname\relax
  \providecommand{\doi}[1]{doi: #1}\else
  \providecommand{\doi}{doi: \begingroup \urlstyle{rm}\Url}\fi

\bibitem[Nogueira et~al.(2019)Nogueira, Yang, Lin, and Cho]{nogueira2019DE}
Rodrigo Nogueira, Wei Yang, Jimmy Lin, and Kyunghyun Cho.
\newblock Document expansion by query prediction, 2019.

\bibitem[Tao et~al.(2006)Tao, Wang, Mei, and Zhai]{DE_for_LM}
Tao Tao, Xuanhui Wang, Qiaozhu Mei, and ChengXiang Zhai.
\newblock Language model information retrieval with document expansion.
\newblock In \emph{Proceedings of the Main Conference on Human Language
  Technology Conference of the North American Chapter of the Association of
  Computational Linguistics}, HLT-NAACL '06, page 407–414, USA, 2006.
  Association for Computational Linguistics.
\newblock \doi{10.3115/1220835.1220887}.
\newblock URL \url{https://doi.org/10.3115/1220835.1220887}.

\bibitem[Efron et~al.(2012)Efron, Organisciak, and Fenlon]{efron12}
Miles Efron, Peter Organisciak, and Katrina Fenlon.
\newblock Improving retrieval of short texts through document expansion.
\newblock In \emph{Proceedings of the 35th International ACM SIGIR Conference
  on Research and Development in Information Retrieval}, SIGIR '12, page
  911–920, New York, NY, USA, 2012. Association for Computing Machinery.
\newblock ISBN 9781450314725.
\newblock \doi{10.1145/2348283.2348405}.
\newblock URL \url{https://doi.org/10.1145/2348283.2348405}.

\bibitem[Liu and Croft(2004)]{croft04}
Xiaoyong Liu and W.~Bruce Croft.
\newblock Cluster-based retrieval using language models.
\newblock In \emph{Proceedings of the 27th Annual International ACM SIGIR
  Conference on Research and Development in Information Retrieval}, SIGIR '04,
  page 186–193, New York, NY, USA, 2004. Association for Computing Machinery.
\newblock ISBN 1581138814.
\newblock \doi{10.1145/1008992.1009026}.
\newblock URL \url{https://doi.org/10.1145/1008992.1009026}.

\bibitem[Agirre et~al.(2010)Agirre, Arregi, and Otegi]{DE_with_wordnet}
Eneko Agirre, Xabier Arregi, and Arantxa Otegi.
\newblock Document expansion based on wordnet for robust ir.
\newblock In \emph{COLING}, 2010.

\bibitem[Sherman and Efron(2017)]{DE_with_external_collection}
Garrick Sherman and Miles Efron.
\newblock Document expansion using external collections.
\newblock In \emph{Proceedings of the 40th International ACM SIGIR Conference
  on Research and Development in Information Retrieval}, SIGIR '17, page
  1045–1048, New York, NY, USA, 2017. Association for Computing Machinery.
\newblock ISBN 9781450350228.
\newblock \doi{10.1145/3077136.3080716}.
\newblock URL \url{https://doi.org/10.1145/3077136.3080716}.

\bibitem[Wolf et~al.(2020)Wolf, Debut, Sanh, Chaumond, Delangue, Moi, Cistac,
  Rault, Louf, Funtowicz, Davison, Shleifer, von Platen, Ma, Jernite, Plu, Xu,
  Scao, Gugger, Drame, Lhoest, and Rush]{wolf2020huggingfaces}
Thomas Wolf, Lysandre Debut, Victor Sanh, Julien Chaumond, Clement Delangue,
  Anthony Moi, Pierric Cistac, Tim Rault, Rémi Louf, Morgan Funtowicz, Joe
  Davison, Sam Shleifer, Patrick von Platen, Clara Ma, Yacine Jernite, Julien
  Plu, Canwen Xu, Teven~Le Scao, Sylvain Gugger, Mariama Drame, Quentin Lhoest,
  and Alexander~M. Rush.
\newblock Huggingface's transformers: State-of-the-art natural language
  processing, 2020.

\bibitem[Bajaj et~al.(2018)Bajaj, Campos, Craswell, Deng, Gao, Liu, Majumder,
  McNamara, Mitra, Nguyen, Rosenberg, Song, Stoica, Tiwary, and
  Wang]{bajaj2018ms}
Payal Bajaj, Daniel Campos, Nick Craswell, Li~Deng, Jianfeng Gao, Xiaodong Liu,
  Rangan Majumder, Andrew McNamara, Bhaskar Mitra, Tri Nguyen, Mir Rosenberg,
  Xia Song, Alina Stoica, Saurabh Tiwary, and Tong Wang.
\newblock Ms marco: A human generated machine reading comprehension dataset,
  2018.

\bibitem[Hofst\"{a}tter et~al.(2020)Hofst\"{a}tter, Zamani, Mitra, Craswell,
  and Hanbury]{long_text_retrieval}
Sebastian Hofst\"{a}tter, Hamed Zamani, Bhaskar Mitra, Nick Craswell, and Allan
  Hanbury.
\newblock Local self-attention over long text for efficient document retrieval.
\newblock In \emph{Proceedings of the 43rd International ACM SIGIR Conference
  on Research and Development in Information Retrieval}, SIGIR '20, page
  2021–2024, New York, NY, USA, 2020. Association for Computing Machinery.
\newblock ISBN 9781450380164.
\newblock \doi{10.1145/3397271.3401224}.
\newblock URL \url{https://doi.org/10.1145/3397271.3401224}.

\bibitem[Bendersky and Kurland(2008)]{doc_homogeneity_measure}
Michael Bendersky and Oren Kurland.
\newblock Utilizing passage-based language models for document retrieval.
\newblock In \emph{Proceedings of the IR Research, 30th European Conference on
  Advances in Information Retrieval}, ECIR'08, page 162–174, Berlin,
  Heidelberg, 2008. Springer-Verlag.
\newblock ISBN 3540786457.

\bibitem[Kwiatkowski et~al.(2019)Kwiatkowski, Palomaki, Redfield, Collins,
  Parikh, Alberti, Epstein, Polosukhin, Kelcey, Devlin, Lee, Toutanova, Jones,
  Chang, Dai, Uszkoreit, Le, and Petrov]{natural_questions}
Tom Kwiatkowski, Jennimaria Palomaki, Olivia Redfield, Michael Collins, Ankur
  Parikh, Chris Alberti, Danielle Epstein, Illia Polosukhin, Matthew Kelcey,
  Jacob Devlin, Kenton Lee, Kristina~N. Toutanova, Llion Jones, Ming-Wei Chang,
  Andrew Dai, Jakob Uszkoreit, Quoc Le, and Slav Petrov.
\newblock Natural questions: a benchmark for question answering research.
\newblock \emph{Transactions of the Association of Computational Linguistics},
  2019.

\bibitem[Rajpurkar et~al.(2018)Rajpurkar, Jia, and Liang]{squad_v2}
Pranav Rajpurkar, Robin Jia, and Percy Liang.
\newblock Know what you don{'}t know: Unanswerable questions for {SQ}u{AD}.
\newblock In \emph{Proceedings of the 56th Annual Meeting of the Association
  for Computational Linguistics (Volume 2: Short Papers)}, pages 784--789,
  Melbourne, Australia, July 2018. Association for Computational Linguistics.
\newblock \doi{10.18653/v1/P18-2124}.
\newblock URL \url{https://www.aclweb.org/anthology/P18-2124}.

\bibitem[Lin et~al.(2014)Lin, Wang, Efron, and Sherman]{Lin2014}
J.~Lin, Y.~Wang, M.~Efron, and Garrick Sherman.
\newblock Overview of the trec-2014 microblog track.
\newblock In \emph{TREC}, 2014.

\bibitem[Voorhees(2004)]{robust_04}
E.~Voorhees.
\newblock Overview of the trec 2004 robust track.
\newblock 2004.

\bibitem[Yang et~al.(2017)Yang, Fang, and Lin]{anserini}
Peilin Yang, Hui Fang, and Jimmy Lin.
\newblock Anserini: Enabling the use of lucene for information retrieval
  research.
\newblock In \emph{Proceedings of the 40th International ACM SIGIR Conference
  on Research and Development in Information Retrieval}, SIGIR '17, page
  1253–1256, New York, NY, USA, 2017. Association for Computing Machinery.
\newblock ISBN 9781450350228.
\newblock \doi{10.1145/3077136.3080721}.
\newblock URL \url{https://doi.org/10.1145/3077136.3080721}.

\bibitem[Klein et~al.(2017)Klein, Kim, Deng, Senellart, and
  Rush]{klein-etal-2017-opennmt}
Guillaume Klein, Yoon Kim, Yuntian Deng, Jean Senellart, and Alexander Rush.
\newblock {O}pen{NMT}: Open-source toolkit for neural machine translation.
\newblock In \emph{Proceedings of {ACL} 2017, System Demonstrations}, pages
  67--72, Vancouver, Canada, July 2017. Association for Computational
  Linguistics.
\newblock URL \url{https://www.aclweb.org/anthology/P17-4012}.

\bibitem[Dehghani et~al.(2017)Dehghani, Zamani, Severyn, Kamps, and
  Croft]{NRM_weak_supervision}
Mostafa Dehghani, Hamed Zamani, Aliaksei Severyn, Jaap Kamps, and W.~Bruce
  Croft.
\newblock Neural ranking models with weak supervision.
\newblock In \emph{Proceedings of the 40th International ACM SIGIR Conference
  on Research and Development in Information Retrieval}, SIGIR '17, page
  65–74, New York, NY, USA, 2017. Association for Computing Machinery.
\newblock ISBN 9781450350228.
\newblock \doi{10.1145/3077136.3080832}.
\newblock URL \url{https://doi.org/10.1145/3077136.3080832}.

\bibitem[Yilmaz et~al.(2019)Yilmaz, Yang, Zhang, and Lin]{Yilmaz2019}
Zeynep~Akkalyoncu Yilmaz, W.~Yang, Haotian Zhang, and Jimmy Lin.
\newblock Cross-domain modeling of sentence-level evidence for document
  retrieval.
\newblock In \emph{EMNLP/IJCNLP}, 2019.

\bibitem[Craswell et~al.(2020)Craswell, Mitra, Yilmaz, Campos, and
  Voorhees]{trec_dl_2019}
Nick Craswell, Bhaskar Mitra, Emine Yilmaz, Daniel Campos, and Ellen~M.
  Voorhees.
\newblock Overview of the trec 2019 deep learning track, 2020.

\bibitem[Abdul-jaleel et~al.(2004)Abdul-jaleel, Allan, Croft, Diaz, Larkey, Li,
  Smucker, and Wade]{rm3}
Nasreen Abdul-jaleel, James Allan, W.~Bruce Croft, O~Diaz, Leah Larkey, Xiaoyan
  Li, Mark~D. Smucker, and Courtney Wade.
\newblock Umass at trec 2004: Novelty and hard.
\newblock In \emph{In Proceedings of TREC-13}, 2004.

\bibitem[Nogueira and Cho(2019)]{nogueira2019passage}
Rodrigo Nogueira and Kyunghyun Cho.
\newblock Passage re-ranking with bert.
\newblock \emph{arXiv preprint arXiv:1901.04085}, 2019.

\bibitem[Beltagy et~al.(2020)Beltagy, Peters, and Cohan]{beltagy2020longformer}
Iz~Beltagy, Matthew~E. Peters, and Arman Cohan.
\newblock Longformer: The long-document transformer, 2020.

\end{thebibliography}
\end{document}